\documentstyle[12pt,aaspp]{article}
\begin{document}

\title{THE DISTRIBUTION OF GALACTIC DISK STARS IN BAADE'S WINDOW}
\author{B. Paczy\'nski, K. Z. Stanek\altaffilmark{1}}
\affil{Princeton University Observatory, Princeton, NJ 08544--1001}
\altaffiltext{1}{On leave from N. Copernicus Astronomical Center,
Bartycka 18, Warszawa 00-716, Poland}
\author{A. Udalski, M. Szyma\'nski, J. Ka\l u\.zny, M. Kubiak}
\affil{Warsaw University Observatory, Al. Ujazdowskie 4, 00--478 Warszawa,
Poland}
\author{and}
\author{M. Mateo}
\affil{Department of Astronomy, University of Michigan, 821 Dennison
Bldg., Ann Arbor, MI~48109--1090}

\begin{abstract}
The color-magnitude diagram of $ \sim 3 \times 10^5 $ stars
obtained for Baade's Window towards the Galactic bulge
with the OGLE project reveals
a surprisingly narrow main sequence due to galactic disk stars
at a distance of $ d \sim 2 $  kpc, i.e. at the location of
the Sagittarius spiral arm.  A more careful analysis indicates
there is an excess in the number of disk stars by a factor $ \sim 2 $
between us and $ d \sim 2.5 $ kpc and a rapid drop
by a factor $ \sim 10 $ beyond that distance.  It is
unlikely that the observed structure is an artifact of
the interstellar extinction, but careful determination of
the extinction is needed before the structure is firmly established.

The narrow main sequence extends down
to stars as faint as $ M_{_V} \approx 7 $, i.e. it is composed
of old stars.  This is not expected in a
conventional disk model, or in a conventional model of a spiral
structure.  However, a strong concentration of old stars towards
spiral arms has been noticed in some other
galaxies, like M51 (Rix \& Rieke 1993),
with the near infrared surface photometry.

We have also found that the relative distribution of
red clump and red giant stars in the Galactic bulge
implies that there is a relatively young stellar population
present there. Color-magnitude diagram data is accessible over the
computer network with anonymous {\tt ftp}.

\end{abstract}


\section{INTRODUCTION}

The Optical Gravitational Lensing Experiment (OGLE, Udalski et al.~1992,
1993b) is an extensive photometric search for the rare cases of
gravitational microlensing of the Galactic bulge stars
by the Galactic disk stars, brown dwarfs and planets.
It provided a huge data base for the studies of other problems
(Szyma\'nski \& Udalski 1993),
in particular the color-magnitude diagrams (Udalski et al.~1993a).
These revealed an expected population of bulge stars, with its turn-off
point, red giant branch and red clump, but also an unexpected
concentration of stars in the blue part of the color-magnitude diagrams.
This feature may be attributed to foreground disk stars,
which seem to concentrate at a single distance, possibly coinciding
with the Sagittarius spiral arm.
This concentration in the color-magnitude diagram was previously noticed by
Terndrup (1988) and Tyson (1991), and even earlier by
Rodgers, Harding \& Ryan (1986), but it shows up much more clearly
in our data (Udalski et al.~1993a), because of unprecedented coverage
combined with very good seeing.  It seems that the full consequences
for the distribution of the Galactic disk stars were not appreciated
in the past, even though Rodgers et al. (1986) and Terndrup (1988) noticed
that the observed number of disk stars is approximately twice as
large as expected in a standard model, and a possible association with the
spiral arm was briefly mentioned by Terndrup (1988) and by Udalski
et al.~(1993a).  In this paper we present the summary of our
observational results, we confront them with the expectations based on the
standard Galaxy model
(Bahcall \& Soneira 1980, hereafter B\&S; Bahcall 1986), and we make
tentative suggestions about the meaning of the apparent conflict.

\section{THE DATA}

Udalski et al.~(1993a) present color-magnitude diagrams (CMDs) of
14 fields in the direction of the Galactic bulge,
which cover nearly one square degree and contain about $6 \times 10^5$
stars. All observations
were made using the 1 meter Swope telescope at the Las  Campanas Observatory,
operated by the Carnegie Institution of Washington, and a $ 2048 \times 2048 $
pixel Ford/Loral CCD detector with the pixel size 0.44 arcsec covering
$15' \times 15'$ field of view.   In this paper we discuss only the nine
partly overlapping fields in Baade's Window, covering a total area of
$40' \times 40'$ and centered at $b = -3.9^o , ~ l = 1.0^o$.
As an example the CMD  for the central
field (BWC) is shown in Figure 1, together with five straight lines
which provide a schematic representation of the location of disk main
sequence stars at the distance of 0.5, 1, 2, 4, and 8 kpc,
with the interstellar extinction adopted following Arp
(1965, cf. eqs. 3 in this paper, with $ A_{_{V,GC}} = 1.5 $),
and the $ (V-I)_0 $ color calculated with the eq. (11) of this
paper.  Most of the diagram is dominated by the bulge stars, with
the disk stars concentrated mostly near the line corresponding
to $\sim 2$ kpc. The part of the diagram dominated
by the disk stars is shown in Figure 2, displaying all stars from the nine
fields in Baade's Window and the same five main sequence lines, as well
as four solid lines corresponding to the disk main sequence stars of the
absolute visual magnitude and the unreddened $ V-I $ color:
$ [M_{_V},(V-I)_0] = ~ (1.0, ~ 0.0), ~ (3.0, ~ 0.2), ~ (5.0, ~ 0.4),
{}~ $ and $~ (7.0,~ 0.6)$, respectively (cf. section 3).
There are $\sim 16,000$ stars plotted in Figure 2, and we believe the
most of them are in the Galactic disk.

The distribution of stars in Figure 2 is very surprising.  A conventional
B\&S model of the Galactic disk has the number density of stars decreasing
exponentially with the distance from the Galactic plane with the
scale height of $ \sim 300 $ pc, and increasing exponentially
towards the Galactic center with the length scale of $ \sim 3.5 $ kpc.
These conspire to give the number density of stars which is approximately
constant with distance while looking through Baade's Window, i.e.
it implies from geometry that we expect
the number of stars to increase by a factor
of $ \sim 4 $ per magnitude.  This is approximately what is observed
out to the distance of $ \sim 2 $ kpc, but there is a dramatic deficit
in the number of stars at larger distances.
There was evidence for this sequence of the Galactic disk stars in
the CMDs obtained previously for Baade's Window
as well as for other areas -- further away from the Galactic
plane: $ (l,b) \approx (0^o, -25^o) $ (Rodgers et al.~1986),
$ (l,b) \approx (0^o, -8^o) $ (Terndrup 1988) -- and closer  to the Galactic
plane: $ (l,b) \approx (1^o, -2^o) $ (Tyson 1991),
and $ (l,b) \approx (14^o, -1^o) $ (Ortolani, Bica \& Barbuy 1992),
but the sequence shows much more clearly in our data.

The interstellar extinction toward Baade's Window merits some discussion.
There are many estimates of extinction towards the globular cluster
NGC 6522 located near the center of our central field BWC (Arp 1965;
Walker \& Mack 1986; Terndrup 1988; Walker \& Terndrup 1991; and references
therein); their estimates of extinction towards NGC 6522 range from
$ A_{_V} = 1.5 $ all the way to $ A_{_V} = 1.78 $.  We adopt
$ A_{_{V,BWC}} = 1.5 $ as a reasonable value (Arp 1965, Terndrup 1988)
for our initial presentation, but we explore a large range of values in
our discussion (section 4).

We estimate the extinction towards the Galactic bulge in our other
eight fields as follows.  We selected the red giants according
to the following conditions:
\begin{equation}
  1.5 < V-I < 2.4, ~~~~~~ 13.0 < V_{_{V-I}} \equiv V - 2.6 ~ (V-I) < 14.5,
\end{equation}
and the red clump stars according to
\begin{equation}
  1.5 < V-I < 2.4, ~~~~~~ 11.5 < V_{_{V-I}} \equiv V - 2.6 ~ (V-I) < 13.0,
\end{equation}
in all nine fields.  There were about 3,000 red giant stars and 4,000
red clump stars in every field.
The parameter $ V_{_{V-I}} $ has been chosen so
that for any particular star its value is not affected by the unknown
extinction.  The distribution of $ V-I $ colors was determined for
each group in every field.  The forms of the distributions were similar
except for a small shift in the $ V-I $ color caused by the differences
in extinction between the fields.   The results are given in Table 1:
column 1 provides the name of the field, columns 2 and 3 give the
galactic coordinates of the field center, columns 4 and 5 give
the difference $ \Delta (V-I) $ between a given field and BWC as
obtained with the red giants (rg) and the red clump stars (rc),
respectively, column 6 gives the total visual absorption adopted for
each field, $ A_{_{V,GB}} $, and the last column contains the
value of the $ V_{_{V-I}} $ parameter at which the distribution
of the red clump stars has its maximum.  With an rms difference of 0.03 mag,
it is clear that the two measures of $ \Delta (V-I) $ agree with each
other quite well.  The average value of extinction for all 9 fields
covering the OGLE Baade Window is larger then that measured towards
NGC 6522 by $ \langle \Delta A_{_{V,GB}} \rangle = 0.09 $.
A reasonable range for the average value of total extinction is
$ 1.6 \leq \langle \Delta A_{_{V,GB}} \rangle \leq 1.9 $.

While the total extinction towards the Galactic bulge seems to be
fairly well established the variation of extinction with distance
along the line of sight is poorly known.
Following Arp (1965) we adopt the following simple law:
\begin{equation}
\begin{array}{llc}
 A_{_V} = A_{_{V,GB}} \times (d / 2 ~ kpc) & {\rm for} & d < 2 ~ kpc \\
 A_{_V} = A_{_{V,GB}}                      & {\rm for} & d > 2 ~ kpc,
\end{array}
\end{equation}
and $ E_{_{V-I}} = A_{_V}/2.6 $, following Dean, Warren, \& Cousins (1978)
and Walker (1985).

Notice that according to Table 1
the distribution of the $ V_{_{V-I}} $ parameter peaks
at the value of $ 12.36 \pm 0.03 $ in all nine fields, indicating that it
is almost unaffected by the interstellar extinction, as designed.
However, there
is some correlation between $ A_{_{V,GB}} $ and $ V_{_{V-I}} $ as
given in Table 1, indicating that our choice of the ratio
$ A_{_V} / E_{_{V-I}} = 2.6 $ was not perfect.

\section{THE STANDARD MODEL}

We calculated the expected distribution of foreground
Galactic disk stars following the standard B\&S model adopting
disk parameters as given by Bahcall (1986, p. 591).  The number density
of disk stars $ n_{_D} $ per cubic parsec is given as
\begin{equation}
 n_{_D} = n_{_D} (R_0) ~ \exp [-|z|/H(M_{_V})] ~
\exp [-(R_{_{GC}}-R_0)/h] , \end{equation}
where $ R_{_{GC}} $ is the galactocentric distance, z is the distance from
the Galactic plane, $ n_{_D} (R_0) $ is the star number density near the
sun, i.e. at the distance $ R_0 = 8 ~ kpc $  (Holtzman et al.~1993
and references therein) from the Galactic center,
$ H(M_{V}) $ is the scale height of the distribution of stars of absolute
visual magnitude $ M_{_V} $, and $ h ~ = ~ 3.5 ~ kpc $ is the disk scale
length.  The scale height $ H(M_{_V}) $ is given by Bahcall (1986, p. 595)
as
\begin{equation}
\begin{array}{llc}
 H(M_{_V}) = 90 ~ pc            & {\rm for}  & M_{_V} < +2.3 , \\
 H(M_{_V}) = 90 ~ pc + 83.9 ~pc ~ (M_{_V} - 2.3) & {\rm for} &
+2.3 < M_{_V} < +5.1, \\
 H(M_{_V}) = 325 ~  pc           & {\rm for}  & +5.1 < M_{_V} .
\end{array}
\end{equation}

The disk luminosity function adopted by Bahcall is based on Wielen (1974).
For stars with $ M_{_V} < +6 $ it is almost identical with the Luyten
(1968) luminosity function, for which Bahcall and Soneira (1980, p. 77) give
analytical approximation:
\begin{equation}
 \Phi (M_{_V}) = 4 \times 10^{-3} ~ [pc^{-3}] ~
{ 10^{0.04x} \over \left( 1 + 0.1^{0.206x} \right) ^{3.4} } ,
{}~~~~~~~ x \equiv M_{_V}-1.28 .
\end{equation}

Combining eqs. (1-3) we can calculate the number density of stars
as a function of distance as seen through Baade's Window.
In particular, we have the relations
\begin{equation}
 z = d \times \sin |b| \approx 0.068 ~ d , ~~~~~~~  R_{_{GC}} = |d-R_0| ,
\end{equation}
where d is the distance from us, and we made an approximation
$ \cos b \approx 1 $.

Notice that at the center of Baade's Window the eq. (4) can be written as
\begin{equation}
\begin{array}{r}
 n_{_{D,BW}} =
n_{_D} (R_0) ~ \exp [(d/3.5 ~ kpc) ( 1 - 238 ~ pc/H(M_{_V}))] , \\
{\rm for} ~~~ b = -3.9^o , ~~~ d < R_0 = 8 ~ kpc,
\end{array}
\end{equation}
i.e. for stars of $ M_{_V} \approx 4 $ the number density is constant
with distance, for the brighter stars the number density is falling
with distance, and for the fainter stars it is increasing with distance.
The numerical coefficients vary across Baade's Window, but only a little.

The area of Baade's Window covered by the OGLE is $(40')^2$ which is
$ 1.35 \times 10^{-4} ~ rad^2 $.
Therefore, the volume contained between distances
$ d_1 $ and $ d_2 $ can be calculated as
\begin{equation}
 V_{1,2} = 4.5 \times 10^{-5} (d_2^3 - d_1^3) .
\end{equation}

The empirical color-magnitude relation for the Pleiades main
sequence was given by Walker (1985) and Prosser, Stauffer \& Kraft (1991).
To this we apply  a simple analytical fit:
\begin{equation}
 (V-I)_{_{ZAMS}} \approx 0.2 \times ( M_{_V} -1.5 ), ~~~~~~~~
1.5 < M_{_V} < 9.0 ,
\end{equation}
which gives the rms deviation of only 0.05 mag in V-I.
The majority of disk stars are somewhat evolved.  According to Castellani,
Chieffi \& Straniero (1992)
the width of the hydrogen burning main sequence is
$ \Delta M_{_V} \approx 1.0 ~ mag $ at a fixed color.  Therefore,
we adopted the following approximate color-magnitude
relation for the disk stars
\begin{equation}
 (V-I)_{_D} \approx 0.2 \times ( M_{_V} -1.0 ), ~~~~~~~~
1.0 < M_{_V} < 8.5 .
\end{equation}
The apparent magnitude and color are given as
\begin{equation}
 V = M_{_V} + A_{_V}(d) + 5 \log ~ (d / 10 ~ pc ) ,
{}~~~~~~~~ V-I = (V-I)_{_D} +E_{_{V-I}}(d) .
\end{equation}
We crudely allowed for the spread of the main sequence
stars in magnitude by adding
to each value of V a random number with a uniform distribution
in the range from --0.5 to +0.5.

We have built our model as a sum of nine fields with the galactic coordinates
and the value of total extinction listed in Table 1, and the volume within
each field adopted as 1/9 of that given with the eq. (9).
Combining the equations presented in this chapter we used a Monte-Carlo
simulation to make Figure 3, which presents the distribution of
the main sequence disk stars expected to be seen in Baade's Window
field of $ (40')^2 $.  It is very different from what is observed
(cf. Figure 2).  In the standard model most stars are expected to be
at the distance $ d > 3 ~ kpc $, while in fact about 90\% of the  stars are
observed to be at $ d < 3 ~ kpc $.  A possible
explanation of this discrepancy is discussed in the next section.

Notice  that our star counts were done in each of the nine fields
as if there was no overlap, i.e. we did not identify stars visible
in more than one field, hence some were counted more than once.  There
is approximately 15\% overlap between the fields, so we have over-counted
by a factor $ \sim $ 1.15.  On the other hand our counts for stars
brighter than $ I = 18.5 $ are only $ \sim $ 80\% complete (Udalski
et al.~1993a, Table 4, Sample \#1).  The two effects almost cancel
each other, and we make no correction for them.  Our counts are
likely to be accurate to $ \sim $ 10\%, i.e. uncertainties in
stellar counts are much smaller than the effects we describe in this paper.

\section{DISCUSSION}

The first modification of the standard model is suggested by the
recent paper by Kent, Dame \& Fazio (1991) who find that the
scale height of the
near infrared disk light is 247 pc near us, i.e. at $ R_0 = 8 $ kpc,
but it is only 165 pc at the galactocentric distance $ R_{_{GC}} = 5 $
kpc.  We modify the standard B\&S  model by multiplying the scale height
as given with the eqs. (5) by a factor
$ (1 - 0.89 ~ R_{_{GC}} / R_0 ) $:
\begin{equation}
 H(M_{_V},R_{_{GC}}) = H(M_{_V},R_0) \times
\left( 1 - 0.89 ~ { R_{_{GC}} \over R_0 } \right) .
\end{equation}
Retaining all other details of the model unchanged the
new distribution of stars in the CMD
was calculated and it is shown in Figure 4.  A visual impression
is that the new distribution is closer to the data, but it
is still not a good match.

To make a quantitative comparison between various models and the data it
is useful to present the distribution of stars as a function of
a parameter related to the distance modulus.  The distance modulus
can be expressed as
\begin{equation}
 \mu \equiv 5\log ~ (d/10 ~ pc) = V - 5 ~ (V-I) -1.0
+ [ 5 ~ E_{_{V-I}}(d) - A_{_V}(d) ] ,
\end{equation}
(cf. eqs. 11 and 12).  The term $ [ 5 ~ E_{_{V-I}}(d) - A_{_V}(d) ] $
describes the interstellar extinction and it is not directly measurable,
it rather depends on the adopted model.  However, the parameter
\begin{equation}
 \mu_0 \equiv V - 5 ~ (V-I) -1.0 ,
\end{equation}
can be calculated for every star from its measured color and magnitude.
If there was no extinction we would have $ \mu = \mu_0 $.
We shall analyze the distribution of stars as a
function of the parameter $ \mu_0 $.

To minimize the contribution of the numerous bulge stars we
consider only the region of the CMD clearly dominated by
disk stars and selected according to the following three inequalities:
\begin{equation}
 V < 19.0, ~~~~~ V-I < 1.4,  ~~~~~~~ (V-I) + 0.15 ~ (V-19.0) < 1.1 .
\end{equation}
All stars observed in nine fields in Baade's Window that satisfied
the inequalities (16) were counted in bins of $ \Delta \mu_0 = 0.2 $.
The result appears in Figures 5, and 7, where it is compared
to several theoretical models described in detail below.

The distributions
expected in the standard B\&S model (solid line) and the KDF
(Kent et al.~ 1991) model with the changing disk scale height
(dotted line) combined with the Arp-like (1965) extinction law
are shown in Figure 5.  For $ A_{_{V,GB}} = 1.5 $
these are the distributions shown in the
color-magnitude diagrams presented in Figures 3 and 4.  It is clear that
neither model fits the data well, but the variable scale height model
looks somewhat better than the one with the constant scale height.

The most dramatic feature in the observed distribution is the large drop
in the number density of stars around $ \mu _0 \approx 10.7 $.
We introduce the
following measure of this drop:
\begin{equation}
\Delta \log N \equiv \log N_1 - \log N_2,
\end{equation}
where $N_1$ and $N_2$ are the number of stars observed in
the ranges $ 9.9 \leq \mu _0 < 10.5 $ and $ 11.1 \leq \mu _0 < 11.7 $,
respectively.  We made an attempt to account for the large value of
$ \Delta \log N $ by varying the distribution
of the interstellar extinction with distance.
We introduced the following formula
\begin{equation}
\begin{array}{llc}
 A_{_V} = 0                                      & {\rm for} & d < d_1, \\
 A_{_V} = A_{_{V,GB}} \times (d-d_1) / (d_2-d_1) & {\rm for} &
d_1 < d < d_2, \\
 A_{_V} = A_{_{V,GB}}                            & {\rm for} & d_2 < d ,
\end{array}
\end{equation}
where $ A_{_{V,GB}} $, $ d_1 $ and $ d_2 $ are the three adjustable parameters.
For every value of $ A_{_{V,GB}} $ we varied $ d_1 $ and $ d_2 $ so as to
maximize the value of $ \Delta \log N $.  The result is shown in Fig. 6 for
the KDF and B\&S disk models together with the observed value
of $ \Delta \log N $.

We selected two models indicated with large dots in Fig. 6.  These correspond
to $ (A_{_{V,GB}}, d_1, d_2) $ equal to $ (1.9~  mag, ~ 1 ~ kpc, ~ 3 ~ kpc ) $
and $ (2.6~ mag, ~ 1 ~ kpc, ~ 3 ~ kpc ) $ for the KDF and B\&S models,
respectively.  The distributions of $ \log N $ versus $ \mu _0 $ are
shown in Figure 7.  It is clear that even though the new models have a
large drop in the number of stars near $ \mu _0 = 10.7 $ they also have
a very steep rise around $ \mu _0 = 9.5 $, unlike the observed distribution.

There are two outstanding features in the distribution of the Galactic
disk stars as observed in Baade's Window that seem to be impossible
to explain with any of the models.  First, the number of stars
observed in the distance range $ 0 < d < 2.5 $ kpc is approximately
a factor of 2 larger than expected.  This has already been
noticed by Rodgers et al.~(1986) and by Terndrup (1988).
Second, there is a very sharp drop in the number of observed stars
beyond 2.5 kpc.  It is intriguing that this is the distance
at which there is the Sagittarius spiral arm (Oort, Kerr \& Westerhout 1958;
Mihalas \& Binney 1981, p. 248, and references therein).
It is natural to have a concentration of young star to a spiral arm.
However, the most of the stars so clearly observed by the OGLE
to form a narrow main sequence in Figure 2
have the absolute magnitude $ 4 < M_{_V} < 7 $.
Such stars have the main sequence life time of many billion years.
It is reasonable to expect that the star formation rate within the
galactic disk has not increased dramatically over the last $ 10^8 - 10^9 $
years, or so.  Hence, the vast majority of those stars must be very
old.  An alternative would require the star formation rate at
$ d \sim 2 $ kpc to be suppressed over billions of years until the very
recent time.  We consider this alternative to be unreasonable.

It is surprising
to find old stars concentrated to a spiral arm, but it is not
unprecedented: there were reports of a similar phenomenon
in the spiral galaxy M51, and elsewhere, based on the near
infrared surface photometry (Rix \& Rieke 1993, and
references therein).  Ours is the first indication of such
a concentration based on the direct observations of the faint and old
main sequence stars.

We have fully explored two leading models of the Galaxy but
we cannot prove beyond any doubt that there is no other explanation
for the observed distribution of disk stars except for a concentration
in the Sagittarius arm.  We have found that the large drop
in the observed number of stars beyond $ d \approx 2.5 $ kpc
cannot be explained with the total interstellar extinction
$ A_{_V} \approx 1.5 $ mag.  A reasonable drop
could be obtained for $ A_{_V} \approx 1.9 $ mag, and the variable
disk scale height, but such a model creates also a large jump in
the number of stars near $ \mu _0 = 9.5 $, a feature that is not
observed.  In the models with constant disk scale height the total extinction
required to account for the large drop in the number of stars at $ \mu = 10.7 $
is far too large to be acceptable ($ A_{_{V,GB}} = 2.6 $),
and they also create a problem with the jump at $ \mu _0 = 9.5 $.

At this time we can offer no alternative
to our conclusion that there is a dramatic decline in the number
of disk main sequence star on the far side of the Sagittarius
spiral arm.

There is another intriguing feature in the color-magnitude diagram:
there are more red clump stars in the
Galactic bulge than there are red giants as defined with the
inequalities (1) and (2).  This implies that the population cannot
be made of very old stars only, with the age $ \sim 10^{10} $ years,
as in that case there would be many more red giants than red clump stars,
as is the case in the  very old open cluster NGC 6791 (Ka\l u\.zny \&
Udalski 1992). In a somewhat younger population the stars spend more time
in the red clump region and less time in the red giant region, below the
red clump.  This phenomenon is well documented by the color-magnitude
diagrams of intermediate age clusters (Breger 1982; Anthony-Twarog et al.
1990; Paez et al.~1990), as well as by the theoretical models (Castellani
et al.~1992; Schuller et al.~1992). A similar effect: the age dependence
of the ratio of the number of red clump stars to the number of brighter
red giant stars has already been pointed out by Barbaro \& Pigatto (1984).
The presence of an age spread in the Galactic bulge has already been
noticed (cf. Frogel 1988; Rich 1991, 1992; Holtzman 1993 et al.;
and references therein), mostly using the bright red giants or
the main sequence turn-off as the age indicators.   As far as we
know the red clump statistics has never been exploited for this purpose.
A quantitative assessment of the age spread is beyond the scope of
this paper, but the readers can use our color-magnitude diagram
for this purpose (available by {\tt ftp} -- see below).

We have discovered yet another feature in the color-magnitude
diagram that is probably related to a similar feature reported by Ortolani
et al.~(1992) in their field at $ (l,b) \approx (14^o,-1^o) $.
Figure 8 shows a sequence of stars between two dashed lines
that extends  upwards from the bulge red clump.  These lines
correspond to the intrinsic colors $ (V-I)_0 = 0.8 $ and
$ (V-I)_0 = 1.28 $, and the horizontal solid  lines correspond to
$ M_{_V} = 1.2 $ at the distance of  1, 2, 4, and 8 kpc. Arp's (1965)
extinction law is again  adopted.  The parameters
$ [M_{_V},(V-I)_0] = (1.2, ~ 1.28) $ correspond to the peak
density of the stars in the red clump of the Galactic bulge (cf.
Table 1).  It is likely that the sequence of stars present in
Figure 8 between the distances of 1 and 4 kpc is made of the disk
red clump stars.  There are enough of them to notice, but not
enough of them to study their CMD morphology.

The distribution of stars in the color-magnitude diagram as observed by
OGLE in Baade's Window is available over the computer network using
anonymous {\tt ftp} on {\tt astro.princeton.edu}.  Login as {\tt ftp},
use your name as a password.  Change directory to {\tt bp/spiral}.
The file {\tt read.me} contains a list of the necessary files
and instructions how to retrieve the data.

\acknowledgments{We are most grateful to Dr. N. D. Tyson for
his many helpful comments and ideas and his careful reading of the
draft of this paper. We also acknowledge discussions with Drs. J. P. Ostriker
and D. N. Spergel.  We also acknowledge a very important comment by the
referee, which made us look more quantitatively at the effects of
interstellar extinction on the apparent distribution of the stars in the
color - magnitude diagram.
This project was supported with the NSF grants AST 9216494
and AST 9216830 and Polish KBN grants No 2-1173-91-01 and BST438A/93.}

\newpage

\begin{figure}[h]
\begin{center}
{\bf FIGURE CAPTIONS}
\end{center}

\caption{The $V-I$ color-magnitude diagram for stars in the central
of the nine Baade's Window fields of the OGLE experiment
(Udalski et al.~1993a).
The five dashed lines show the approximate location of the Pleiades
main sequence (Walker 1985) at the distance of 0.5, 1, 2, 4, and 8 kpc,
with the interstellar reddening adopted following Arp (1965).
The vast majority of the stars are in the Galactic bulge.  The disk stars
are concentrated along the main sequence line corresponding to the
distance of 2 kpc.}

\caption{The $V-I$ color-magnitude diagram for stars in all
nine Baade's Window fields of the OGLE experiment (Udalski et al.~1993a).
As in Figure 1, the five dashed lines show the approximate location of the
Pleiades main sequence at the distance of 0.5, 1, 2, 4, and 8 kpc, with the
interstellar reddening adopted following Arp (1965).  Also shown are
four solid lines that correspond  to the disk main sequence stars of the
absolute visual magnitude and the unreddened $ V-I $ color:
$ [M_{_V},(V-I)_0] = $ (1.0,0.0), (3.0,0.2), (5.0,0.4), and
(7.0,0.6). Almost all stars are in the Galactic disk, and about 90\% of
them are closer than 3 kpc.}

\caption{The $V-I$ color-magnitude diagram for the Galactic disk
stars as expected according to the standard model (Bahcall 1986,
Bahcal \& Soneira 1980).  This Monte-Carlo simulation was done for
the field of $ (40')^2 $ centered on Baade's Window.  As in Fig. 1
the five dashed lines show the approximate location of the Pleiades
main sequence at the distance of 0.5, 1, 2, 4, and 8 kpc, with the
interstellar reddening adopted following Arp (1965) and described
in section 2.  Also shown are
four solid lines corresponding to the disk main sequence stars of the
absolute visual magnitude and the unreddened $ V-I $ color:
$ [M_{_V},(V-I)_0] = $ (1.0,0.0), (3.0,0.2), (5.0,0.4), and (7.0,0.6).
The most of the stars in this model are beyond 3 kpc.}

\caption{The same as Fig. 3 but for a model with the disk scale height
decreasing towards the Galactic center (Kent et al.~1991, cf. eq. 13).}

\caption{The distribution of the number of Galactic disk stars in Baade's
Window, N, as a function of distance modulus $ \mu_0 $ (cf. eq. 15) is shown.
The error bars correspond to $ N^{-1/2} $, where $ N $ is the number of stars
that satisfy the inequalities (16) per bin of $0.2 ~ mag$.  The solid lines
represent the standard model (B\&S, Bahcall \& Soneira 1980), while the dotted
lines correspond to the model with the disk scale height decreasing
towards the Galactic center (KDF, Kent et al.~1991, cf. eq. 13), with the
interstellar extinction adopted following Arp (1965, cf. eqs. 3 and Table 1).
with $ A_{_{V,GB}} = 1.5 $ and $ A_{_{V,GB}} = 1.9 $, as indicated.}
\end{figure}

\newpage

\begin{figure}
\caption{The variation of the largest value of the
$ \Delta \log N \equiv \log N_1 - \log N_2 $ parameter (cf. eq. 17) with
the total extinction towards the Galactic Bulge $ A_{_{V,GB}} $ is
shown for the KDF (Kent et al. ~1991 ) and B\&S (Bahcall and Soneira 1986)
disk models with the solid and short dash lines, respectively.
The observed value of $ \Delta \log N $ is shown with a horizontal
long dash line.  The distributions corresponding to the two large dots are
shown in Fig. 7.}

\caption{The same as Fig. 5 but corresponding to the two extreme
interstellar extinction laws (cf. eqs. 18 and discussion) indicated
in Fig. 6 with the two large dots.
The KDF (Kent et al. ~1991 ) and B\&S (Bahcall and Soneira 1986)
disk models are shown with the solid and short dash lines, respectively.}

\caption{The $V-I$ color-magnitude diagram for stars in all
nine Baade's Window fields of the OGLE experiment (Udalski et al.~1993a).
The five dashed lines show the approximate location of the
Pleiades main sequence at the distance of 0.5, 1, 2, 4, and 8 kpc, with the
interstellar reddening adopted following Arp (1965).  Also shown are
two dashed lines connected with four horizontal solid lines.
These solid lines  correspond to the location of the red clump stars at the
distance of 1, 2, 4, and 8 kpc.}

\end{figure}

\begin{planotable}{lllllll}
\tablewidth{39pc}
\tablecaption{Parameters for the nine Baade's Window fields}
\tablehead{
\colhead{field}	& \colhead{$l$}	& \colhead{$b$} &
\colhead{$\Delta (V-I)_{rg}$} 	& \colhead{$\Delta (V-I)_{rc}$}	&
\colhead{$A_{_{V,GB}}$}		& \colhead{$V_{_{V-I}}$} }

\startdata
BW1 	& 1.09 	& $-$3.60 & $-$0.04  & $+$0.01  & 1.46 & 12.42 \nl
BW2	& 0.71 	& $-$3.81 & $+$0.07  & $+$0.07  & 1.68 & 12.37 \nl
BW3 	& 0.92 	& $-$4.19 & $+$0.02  & $+$0.07  & 1.62 & 12.38 \nl
BW4	& 1.30 	& $-$3.98 & $+$0.01  & $+$0.00  & 1.51 & 12.32 \nl
BW5	& 0.90 	& $-$3.71 & $+$0.12  & $+$0.11  & 1.80 & 12.35 \nl
BW6	& 0.82	& $-$4.00 & $+$0.06  & $+$0.09  & 1.70 & 12.32 \nl
BW7	& 1.11 	& $-$4.08 & $+$0.05  & $+$0.02  & 1.59 & 12.34 \nl
BW8	& 1.20 	& $-$3.79 & $-$0.02  & $-$0.04  & 1.42 & 12.37 \nl
BWC	& 1.01 	& $-$3.89 & $+$0.00  & $+$0.00  & 1.50 & 12.37
\end{planotable}

\end{document}